\documentclass[aps,prl,reprint,groupedaddress,showpacs,showkeys,floatfix,longbibliography]{revtex4-1}
\usepackage{watermark}
 \usepackage{amsmath}
 \usepackage{amsfonts}
 \usepackage{graphicx}
 \usepackage{epsfig}
 \usepackage{epstopdf}
 \usepackage{comment}

 \usepackage{hyperref}
 \hypersetup{
    bookmarks=true,         
    unicode=false,          
    pdftoolbar=true,        
    pdfmenubar=true,        
    pdffitwindow=false,     
    pdfstartview={FitH},    
    pdftitle={Positron cooling and annihilation in noble gas atoms},   
    pdfauthor={D. G. Green},     
    pdfsubject={Positron_hlikeions},   
    pdfcreator={D. G. Green},   
    pdfproducer={Producer}, 
    pdfkeywords={keywords}, 
    pdfnewwindow=true,      
    colorlinks=true,       
    linkcolor=blue,          
    citecolor=blue,        
    filecolor=magenta,      
    urlcolor=blue           
}

\begin{document}
\title{Positron cooling and annihilation in noble gases}
\author{D.~G. Green}
\email[Correspondences to:~]{d.green@qub.ac.uk}
\affiliation{
Centre for Theoretical Atomic, Molecular and Optical Physics,
School of Mathematics and Physics,\\
Queen's University Belfast, Belfast, BT7\,1NN, Northern Ireland, United Kingdom}
\date{\today}

\maketitle

{\bf 
Understanding the dynamics of positron cooling in gases, including the fraction of positrons surviving to thermalisation, is critical for accurate interpretation of positron lifetime spectra, for the development of efficient positron cooling in traps and accumulators, and for a cryogenically cooled, ultra-high-energy-resolution, trap-based positron beam. Here, positron cooling and annihilation in noble gases is simulated using accurate scattering and annihilation cross sections calculated \emph{ab initio} with many-body theory. 
It is shown that a strikingly small fraction of positrons survive to thermalisation: 
$\sim$0.1 in He, $\sim$0 in Ne (due to cooling effectively stalling in the relatively deep momentum-transfer cross-section minimum), $\sim$0.15 in Ar, $\sim$0.05 in Kr and $\sim$0.01 in Xe. 
For Xe, the time-varying annihilation rate $\bar{Z}_{\rm eff}(\tau)$ is shown to be highly sensitive to the depletion of the distribution due to annihilation, conclusively explaining the long-standing discrepancy between gas-cell and trap-based measurements.
The \emph{ab initio} calculations enable the first simultaneous probing of the energy dependence of the scattering cross section and annihilation rate.
Overall, the use of the accurate atomic data gives $\bar{Z}_{\rm eff}(\tau)$ in close agreement with experiment for all noble gases except Ne, the experiment for which is proffered to have suffered from incomplete knowledge of the fraction of positrons surviving to thermalisation and/or the presence of impurities.
}

Observation of lifetime spectra for positrons annihilating in gases was one of the first sources of information on positron interaction with atoms and molecules (see e.g., \cite{Falk:1964,Tao:1964,Paul:1964,Coleman:1975} and \cite{Griffith:1979, Charlton:1985} for reviews). 
In particular, measurements of the time-varying normalised annihilation rate $\bar{Z}_{\rm eff}(t)$ during positron thermalisation provided information on the energy dependence of the scattering and annihilation cross sections. 
Understanding the dynamics of positron cooling, including the fraction of positrons surviving to thermalisation, is critical for accurate interpretation of the positron lifetime experiments. 
Incomplete thermalisation was suspected to be responsible for the lack of consensus among the $\bar{Z}_{\rm eff}$ data in Xe \cite{Wright:1985}, while modelling of $\bar{Z}_{\rm eff}(t)$ \cite{Campeanu:1981,Campeanu:1982} revealed deficiencies in the theoretical data for neon and the heavier noble-gas atoms. Understanding of positron kinetics is also crucial for the development of efficient positron cooling in traps and accumulators \cite{AlQaradawi:2000}, and for a cryogenically cooled, ultra-high-energy-resolution, trap-based positron beam \cite{Natisin:2014,Natisin:2016}. 

Despite the importance of long-standing positron-cooling experimental results \cite{Griffith:1979, Charlton:1985}, there has been a paucity of theoretical studies of positron cooling in gases.
Previous studies have mainly employed the diffusion or Fokker-Planck (FP) equation \cite{Orth:1969,Campeanu:1977,Campeanu:1981,Campeanu:1982,Shizgal:1987,Boyle:2014}. They used semi-empirical or model cross sections, e.g., calculated in the polarised-orbital approximation \cite{polorbital_he,polorbital_he2,McEachran:Ne:1978,arphase,krxephase}, 
yielding limited success in describing the experiments.

Recently, many-body theory (MBT) was used to provide an accurate and essentially complete description of low-energy positron interactions with noble-gas atoms, taking full account, \emph{ab initio}, of the strong positron-atom and electron-positron correlations, including virtual-positronium formation \cite{PhysScripta.46.248,dzuba_mbt_noblegas,DGG_posnobles,DGG:2015:core}. It yielded excellent agreement between theory and experiment for the scattering cross sections, annihilation rates \cite{DGG_posnobles,DGG:2015:core}, and $\gamma$-spectra \cite{DGG:2015:core}, firmly establishing the relative contributions of annihilation on various atomic orbitals. It has also been successfully applied to positron scattering and annihilation on hydrogenlike ions \cite{DGG_hlike} and to study the effect of positron-atom correlations on positron molecule $\gamma$ spectra \cite{DGG_molgamma,DGG_molgammashort}.

Here, we show that the MBT data enables accurate modeling of positron cooling and annihilation in noble gases. Using the MBT data in Monte-Carlo (MC) simulations \cite{Farazdel:1977,DGG_anticool}, we calculate the time-evolving positron momentum distribution $f(k,\tau)$ 
[we work in units where $\tau$ is the time (in ns) scaled by the number density of the gas $n_g$ (in amagat):  $\tau=n_gt$] and from this the time-varying annihilation rate $\bar{Z}_{\rm eff}(\tau) = \int Z_{\rm eff}(k) f(k,\tau) dk/\int f(k,\tau) dk$, which can be compared with experiments. 
The fraction of positrons surviving to thermalisation is shown to be strikingly small. 
$\bar{Z}_{\rm eff}(\tau)$ is shown to be sensitive to the amount of particles annihilating before thermalisation, conclusively explaining the long-standing discrepancy between the gas-cell and trap-based $\bar{Z}_{\rm eff}$ results in Xe. 
Importantly, the approach enables the simultaneous probing of the energy dependence of the cross section and $Z_{\rm eff}$. Overall, the MBT-based MC calculations give the best agreement with experiment for all the noble gases to date.

\begin{figure}[t!!]
\begin{center}
\includegraphics[width=0.45\textwidth]{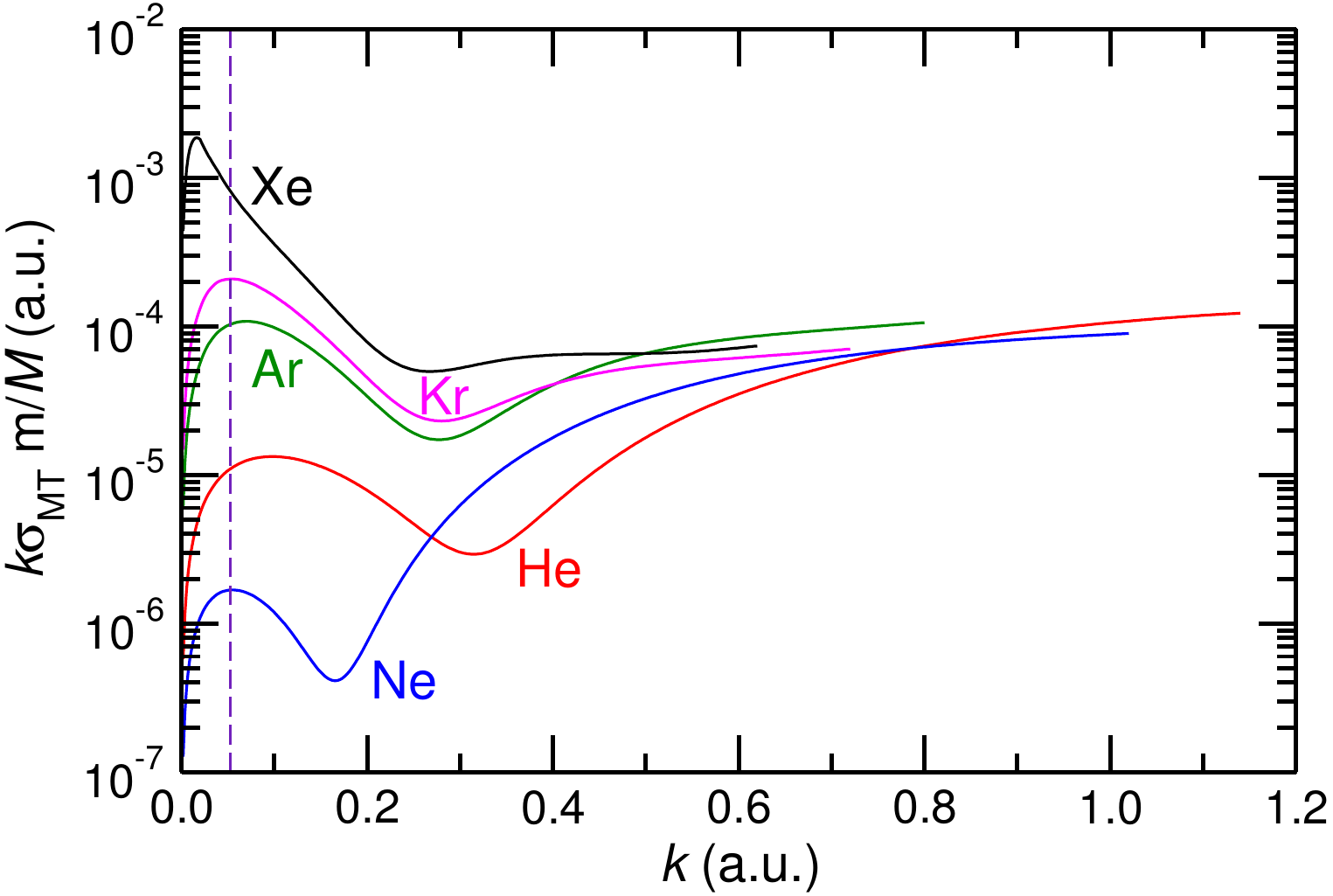}
\caption{The momentum-diffusion coefficient $B(k)/k_{\rm B}T = k\sigma_{\rm t}(k)m/M$.
Thermal positron momentum $k_{\rm th}=\sqrt{3k_{\rm B}T}\sim0.0528$ a.u.~at $T=293$ K (vertical dashed line).
\label{fig:kommtxsec}}
\end{center}
\end{figure}
\section{Results}

\subsection{Examination of the key physical quantities governing positron cooling and annihilation}
At energies below the positronium-formation threshold, energy loss in atomic gases proceeds only via elastic collisions through momentum transfer. 
The process of positron thermalisation in a Maxwellian gas of temperature $T$ is governed by the mean squared change in momentum per unit time $\langle \Delta k^2/\delta \tau \rangle = 2B(k)$, where $B(k)\equiv k\sigma_{\rm t}(k)k_{\rm B}T m/M$ \cite{physkin}, $k$ and $m$ are the positron momentum and mass, $M$ is the mass of the gas atom and $\sigma_{\rm t}$ is the positron-atom momentum-transfer cross section. It is calculated as $\sigma_{\rm t}=4\pi k^{-2}\sum_{\ell=0}^{\infty}(\ell+1) \sin^2(\delta_{\ell} - \delta_{\ell+1})$ \cite{quantummechanics}, where $\delta_{\ell}(k)$ is the scattering phase shift for a positron of angular momentum $\ell$. 
Figure~\ref{fig:kommtxsec} shows $B(k)$ for He--Xe, calculated using phase shifts for $\ell$=0, 1 and 2 from MBT \cite{DGG_posnobles}, with $\ell>2$ partial waves described by the leading $k^2$ term in the expansion \cite{omalley}. (Higher-order terms in the phaseshifts for $\ell>2$ \cite{AliFraser} contribute negligibly to $\sigma_{\rm t}$, e.g., $<3\%$ in Xe at the Ps-formation threshold,  quickly reducing to $<1\%$ at $k<0.55$ a.u., and can thus be neglected.)
$B(k)$ exhibits a minimum for all atoms in the sequence He--Xe, which becomes less pronounced as one moves through the sequence (with the exception of Ne, which has the deepest minimum). 
As we will see, this leads to ``trapping" of positrons and slowing down of the cooling process in this momentum range.

The annihilation cross section in many-electron targets of number density $n_{\rm g}$ is parametrised as 
$\sigma_{\rm a}= \pi r_0^2 Z_{\rm eff} {c}/{v_r}$ \cite{Fraser,Pomeranchuk},
where $r_0$ is the classical electron radius, $c$ is the speed of light, $v_r$ is the positron speed relative to the target, and $Z_{\rm eff}$ is the dimensionless effective number of electrons that contribute to the annihilation process. It is defined as $Z_{\rm eff} = \lambda/(\pi r_0^2cn_g)$ where $\lambda$ is the true annihilation rate and $\pi r_0^2c$ is the free electron-positron annihilation rate. 
Positron-atom and positron-electron correlations can result in $Z_{\rm eff}$ being greater than the actual number of valence electrons on which positrons predominantly annihilate  \cite{PhysScripta.46.248,dzuba_mbt_noblegas,arphase,krxephase,DGG_posnobles}. 
$Z_{\rm eff}(k)$ has been calculated recently for the noble gases via diagrammatic MBT for s, p and d-wave positrons, taking full account of correlations \cite{DGG_posnobles}. 
The s, p and d-waves provide sufficient accuracy to model positron cooling below the Ps-formation threshold: although the contribution of f-wave positrons to the total annihilation rate is expected to be $\sim 10\%$ near the Ps-formation threshold for Xe, we will see that annihilation is unimportant until the positrons have cooled to close to the minimum in $B(k)$, at which point the f-wave contribution is expected to be negligible ($\sim 1\%$: we note that $Z_{\rm eff}(k)\sim k^{2\ell}$ as as $k\to0$ \cite{DGG_posnobles}). 
From He to Xe, $Z_{\rm eff}$ becomes increasingly large and strongly peaked at low momenta (see Fig.~2, calculated including annihilation on core and valence electrons). This effect is due to the existence of positron-atom virtual levels \cite{Goldanski}, signified by large scattering lengths for Ar, Kr and Xe (see Table I in \cite{DGG_posnobles}).

\begin{figure}[t!]
\begin{center}
\includegraphics[width=0.4\textwidth]{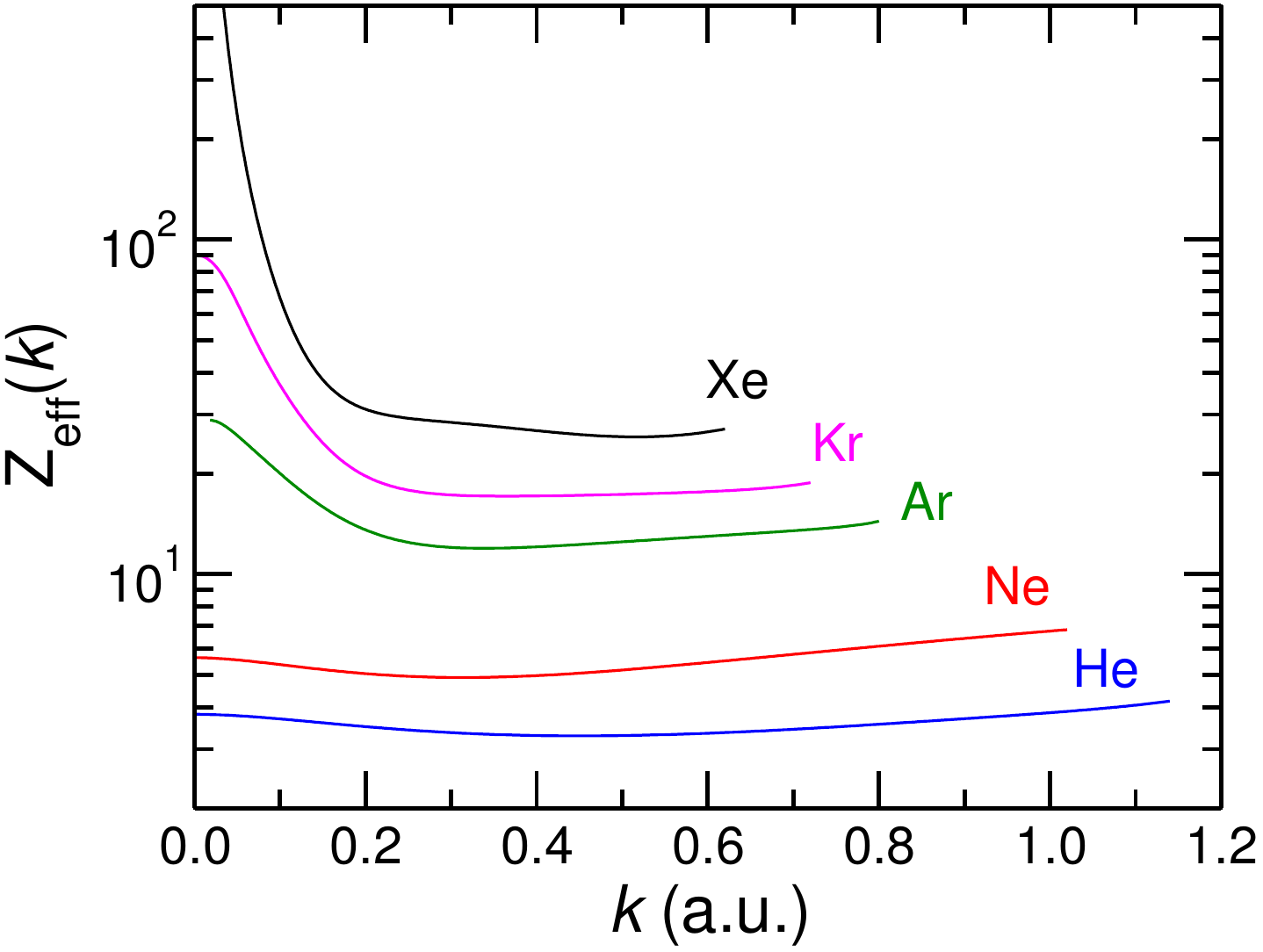}\\
\caption{$Z_{\rm eff}(k)$ vs positron momentum $k$ calculated using MBT including s, p and d-wave positrons annihilating on valence $n$ and inner valence $(n-1)$ subshells of the noble gases.\label{fig:zeffk}}
\end{center}
\end{figure}

\subsection{Simulations of positron cooling and annihilation\label{sec:results}}
\begin{figure}[t!!]
\begin{center}
\includegraphics[width=0.48\textwidth]{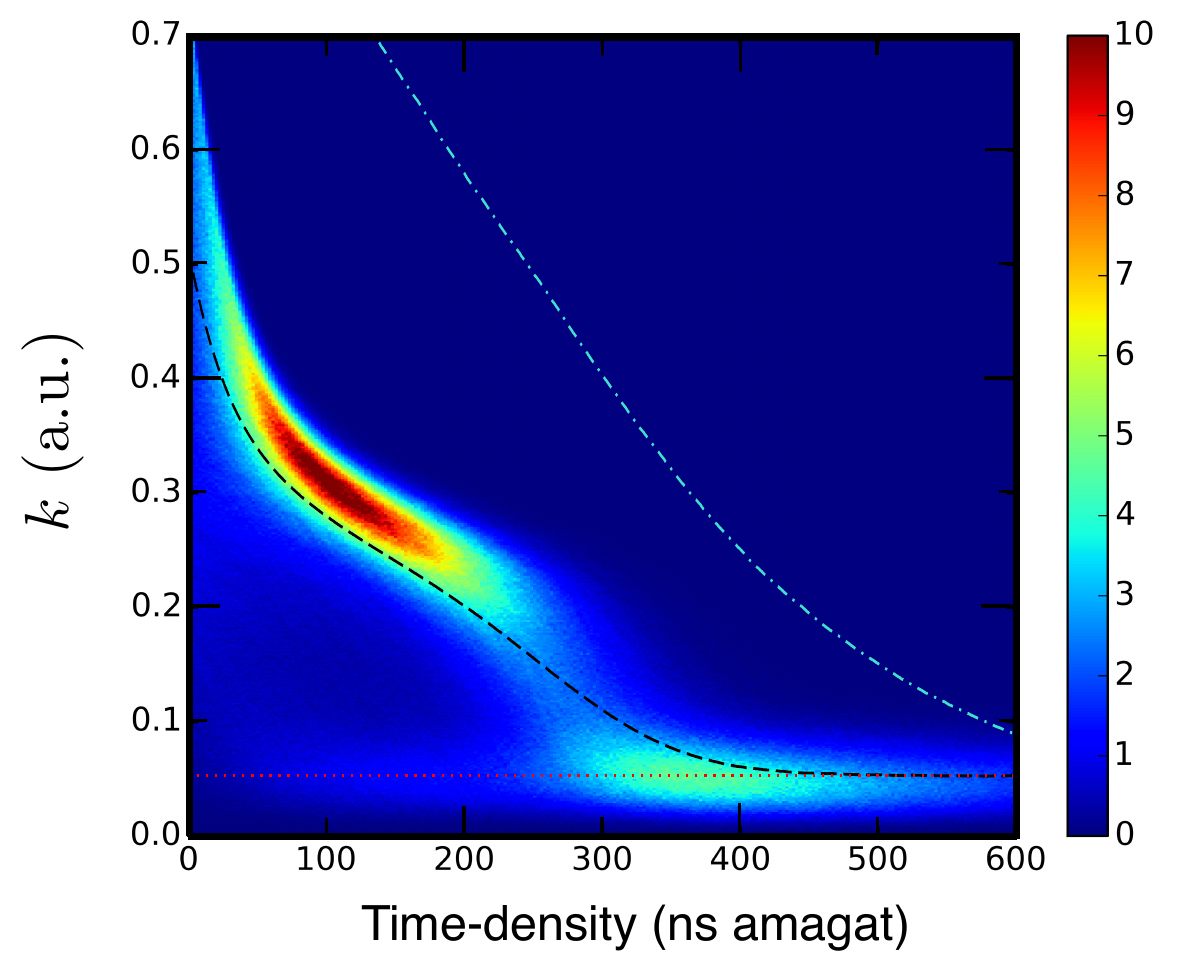}
\caption{Density plot of positron momentum distribution $f(k,\tau)$  for Ar, normalised as $\int f(k,\tau)\,dk=F(\tau)$, the fraction of positrons surviving (dashed-dotted line), calculated using 50,000 positrons initially distributed uniformly in energy up to the Ps-formation threshold. Also shown is the r.m.s.~momentum (black dashed line) and thermal momentum $k_{\rm th}=\sqrt{3k_{\rm B}T}\sim 0.0528$ a.u.~at $T=293$ K (dotted line).
\label{fig:fk}}
\end{center}
\end{figure}

The time-evolving momentum distribution $f(k,\tau)$ for positrons cooling and annihilating in a noble gase is determined via Monte-Carlo simulations based on the MBT scattering and annihilation cross sections (see `Methods' for details on the Monte-Carlo algorithm). 
Figure~\ref{fig:fk} is an example of $f(k,\tau)$, calculated for Ar for 50,000 positrons initially distributed uniformly in energy up to positronium-formation threshold. [Also see Supplemental Material for corresponding  \href{http://www.am.qub.ac.uk/users/dgreen09/coolingvideos.html}{videos} \footnote{\href{http://www.am.qub.ac.uk/users/dgreen09/coolingvideos.html}{http://www.am.qub.ac.uk/users/dgreen09/coolingvideos.html}} of the evolution of $f(k,\tau)$ for of all the noble gases.]
The initial distribution is seen to quickly evolve ($\lesssim$100 ns amg) to a strong Gaussian-like peak near the minimum in $B(k)$, $k_{\rm min}\sim 0.3$ a.u, which then evolves rather slowly, producing the knee-like feature in the figure.
This bunching effect becomes less effective as one moves through the sequence from He to Xe since the minimum in $B(k)$ becomes less pronounced (with the exception of Neon, which has the deepest minimum). However, even in Xe the formation of a peak in $f(k,\tau)$ at $k_{\rm min}\sim 0.25$ a.u. at small time-densities is still evident. 
Since for any realistic initial distribution most positrons will have initial momenta $k> k_{\rm min}$, such bunching should be expected, making the overall cooling times somewhat insensitive to the exact form of the initial distribution (see below for further details).
The small number of positrons with initial momenta below the minimum at early times cool to thermal energies relatively unimpeded, and they form a second, much smaller peak around thermal momentum $k_{\rm th}\sim \sqrt{3k_{\rm B}T}\sim0.0528$ a.u.
As the time-density increases, the Gaussian-like part of the distribution traverses the minimum, roughly maintaining its form as it does so. 
As more positrons cool below the minimum, the two peaks in the distribution merge, eventually evolving towards the Maxwell-Boltzmann distribution.

\begin{figure}[t!]
\begin{center}
\includegraphics[width=0.42\textwidth]{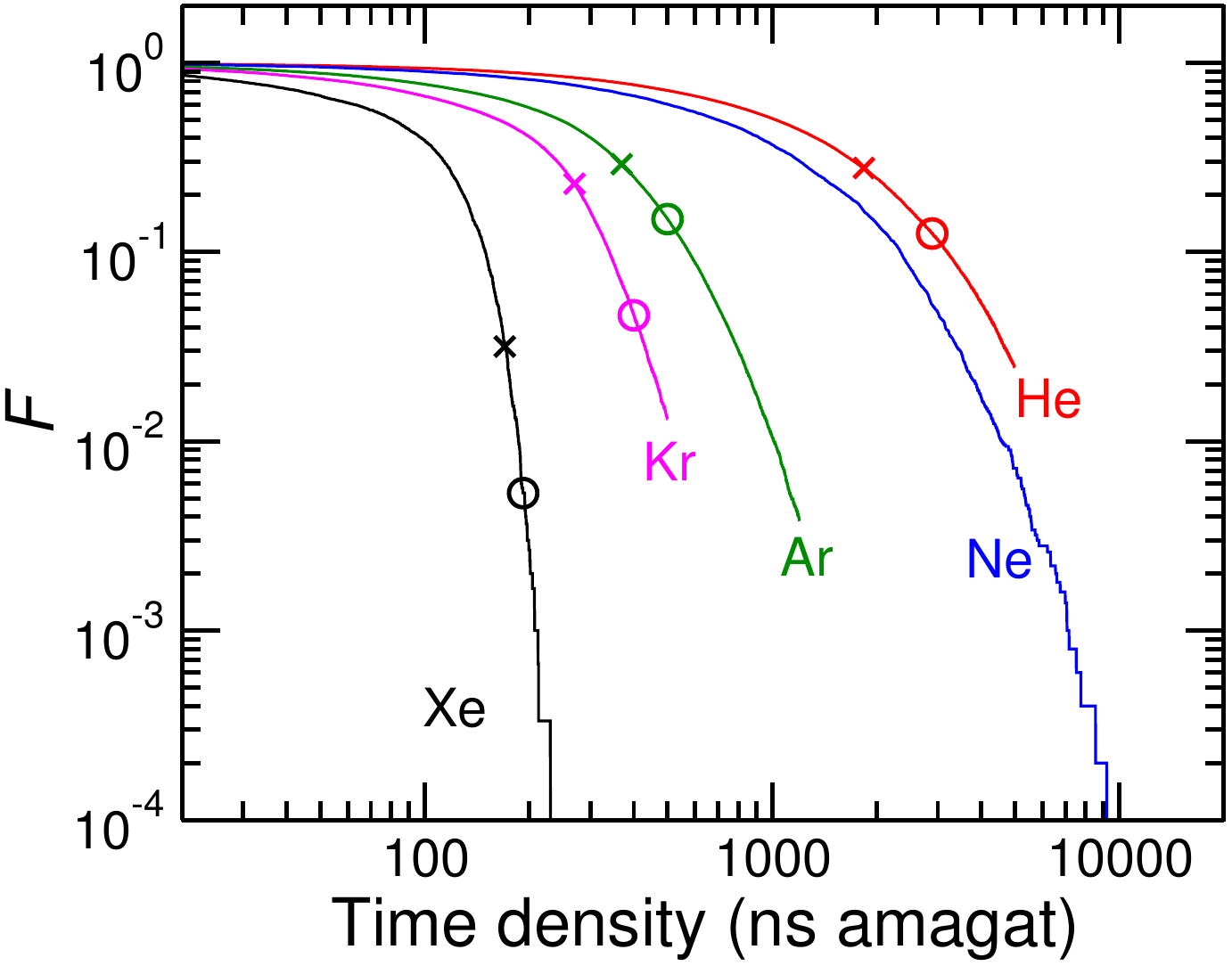}~~~~~~~
\caption{Calculated fraction of positrons remaining vs.~time-density. Crosses indicate the $\bar{Z}_{\rm eff}$ shoulder lengths (defined in the text); circles indicate the time-density of complete thermalisation, at which the r.m.s.~momentum of the distribution $k_{\rm rms}$ is within 1\% of $k_{\rm RT}=\sqrt{3k_BT}\sim 0.0527$ a.u., the momentum of a positron in thermal equilibrium at 293 K. \label{fig:annfrac}}
\end{center}
\end{figure}

As seen in Fig.~\ref{fig:fk}, only a small fraction of positrons survive before the distribution thermalizes at $\tau \sim$ 400 ns amagat. This fraction is even smaller in other noble gases (see Fig. 3), which has important consequences for the interpretation of measured lifetime spectra (see below).
We define the `complete' thermalisation time $\tau_{\rm th}$ as the time-density at which the r.m.s.~momentum of the distribution is within 1\% of $k_{\rm th}\sim \sqrt{3k_{\rm B}T}$, the r.m.s.~for a Maxwell-Boltzmann distribution at 293 K. 
The values of $\tau_{\rm th}$ are marked in Fig.~\ref{fig:annfrac}, and are presented in Table \ref{table:times}, alongside the fractions of initial positrons remaining at that time-density $F(\tau_{\rm th})$. 
This fraction is a mere $F(\tau_{\rm th})=0.11$ for helium, and reduces by more than an order of magnitude for xenon.
Perhaps most remarkably, the fraction of positrons surviving to thermalisation in Ne is zero. 
In neon, cooling effectively stalls at the minimum of $B(k)$ (see Fig.~\ref{fig:kommtxsec}), with positrons eventually succumbing to annihilation (in spite of a relatively small $Z_{\rm eff}\sim 6$) before they can cool further.

\begin{table}[t!]
\caption{Complete thermalisation time-density $\tau_{\rm th}$ and $\bar Z_{\rm eff}$ shoulder lengths $\tau_{\rm s}$ (both defined in the text) in units of ns amagat, and fraction $F$ of positrons remaining at those times. \label{table:times}}
\begin{center}
\begin{tabular}{lccccc}
\hline
\hline
&He & Ne & Ar & Kr & Xe \\
\hline
$\tau_{\rm th}$\footnote{This work, initial uniform energy distribution, including depletion of particles due to annihilation, except for Ne and Xe, for which the distribution calculated neglecting annihilation was used. 
} 
 & 2890$\pm30$ & 21000$\pm50$ & 500$\pm$5 & 400$\pm$10 & 192$\pm$10 \\
$\tau_{\rm th}$\footnote{This work, initial positron energies equal to the positronium-formation threshold energy,  including depletion of particles due to annihilation, except for Ne and Xe, for which the distribution calculated neglecting annihilation was used. } & 3035$\pm 30$  &$23000\pm 500$ & 505$\pm 5$ & 460$\pm 10$ & 240$\pm 10$ \\
\hline
$\tau_{\rm s}$\footnotemark[1]
& 1839 & 12075 &369 & 309 & 170$\pm 15$ \\
$\tau_{\rm s}$\footnotemark[2]
& 1885  & 12300  & 382  & 440$\pm50$  & 185$\pm 20$  \\
$\tau_{\rm s}$\footnote{Experiment of Coleman \emph{et al.}~\cite{Coleman:1975}.} &$1700\pm50$& $2300\pm200$ &$362\pm5$&  $296\pm6$ & $200\pm20$ \\
$\tau_{\rm s}$\footnote{Experiment of Coleman \emph{et al.}, determined via intersection of straight line fits to the lifetime spectra \cite{Coleman:1975}.} 
&--& 1700$\pm200$ &--& --& -- \\
$\tau_{\rm s}$\footnote{Experiment of Wright \emph{et al.}~\cite{Wright:1985}.} & -- &-- &--& $325\pm6$& $178\pm 3$ \\
$\tau_{\rm s}$\footnote{Fokker-Planck calculation of Campeanu and Humberston \cite{Campeanu:1977}, and Campeanu \cite{Campeanu:1981,Campeanu:1982} using polarised-orbital cross sections of McEachran \emph{et al.}~\cite{polorbital_he,polorbital_he2,McEachran:Ne:1978,arphase,krxephase}. } & 1878 &$>$5000 &311 & 288 &130 \\
$\tau_{\rm s}$\footnote{Diffusion equation calculation using model potential and enhancement-factor modified zeroth-order annihilation rate \cite{Boyle:2014}.} & 1618 & -- & -- &-- &-- \\
\hline 
$F(\tau_{\rm th})$\footnotemark[1]
&0.11 & 0.00 & 0.15& 0.046 & 0.0069\\
$F(\tau_s)$\footnotemark[1]
& 0.28 & 0.00 &0.29 & 0.23 & 0.0310\\
\hline
\hline
\end{tabular}
\end{center}
\label{default}
\end{table}%

\subsection{Comparison with experiment: time-varying annihilation rate parameter $\bar{Z}_{\rm eff}$.}

\begin{figure*}[t!!]
\begin{center}
\includegraphics[width=0.93\textwidth]{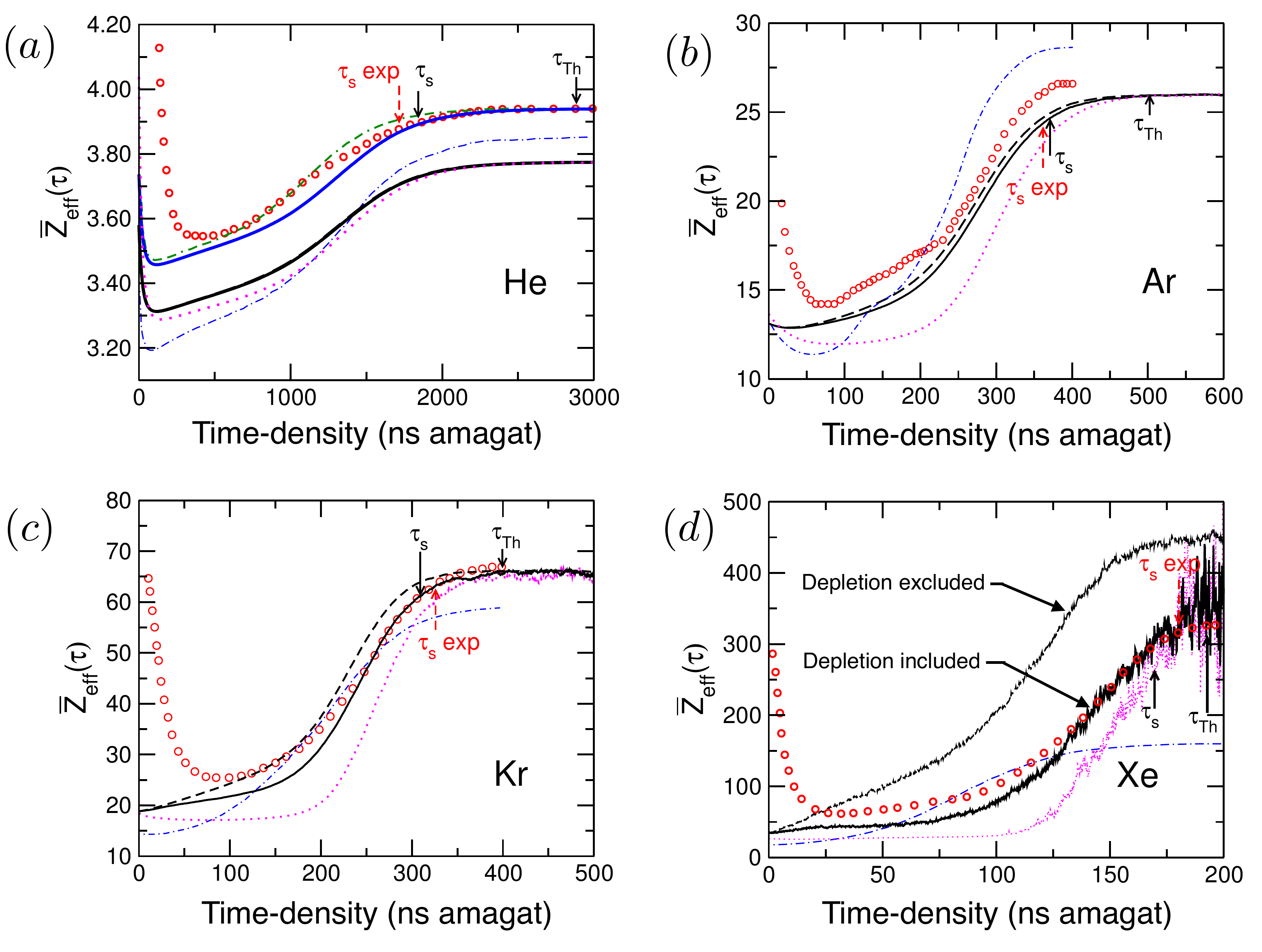}
\caption{$\bar{Z}_{\rm eff}(\tau)$ for He, Ar, Kr and Xe calculated using $f(k,\tau)$ excluding and including loss of particles due to annihilation, initially distributed uniformly in energy (black dashed and solid lines, respectively), and including annihilation with initial energy equal to the Ps-formation threshold (magenta dotted line). 
Also shown for He: experiment of Coleman \emph{et al.} \cite{Coleman:1975} (red circles), present calculation with $\bar{Z}_{\rm eff}$ scaled to the measured value of $\bar{Z}_{\rm eff}$=3.94 (blue solid line), FP calculation of Campeanu \cite{Campeanu:1977}
(blue dashed-dotted line) and model calculations of Boyle \emph{et al}.~\cite{Boyle:2014} scaled to $\bar{Z}_{\rm eff}=3.94$ (green dash-dash-dotted line);
Ar: experiment of Coleman \emph{et al.}~\cite{Coleman:1975,Campeanu:1981} (red circles)
and FP calculation of Campeanu \cite{Campeanu:1981};
Kr and Xe: experiment of Wright \emph{et al.} \cite{Wright:1985} (red circles) and FP calculations of Campeanu and Humberston \cite{Campeanu:1977} (blue dashed-dotted lines);
Ne: experiment of Coleman \emph{et al.}~\cite{Coleman:1975} (red circles) and FP calculation of Campeanu and Humberston \cite{Campeanu:1977} (blue dashed-dotted lines). 
Arrows mark the calculated (black) and experimental shoulder lengths (red) $\tau_s$. 
\label{fig:zefft}}
\end{center}
\end{figure*}

\begin{figure*}[t!!]
\begin{center}
\includegraphics[width=0.93\textwidth]{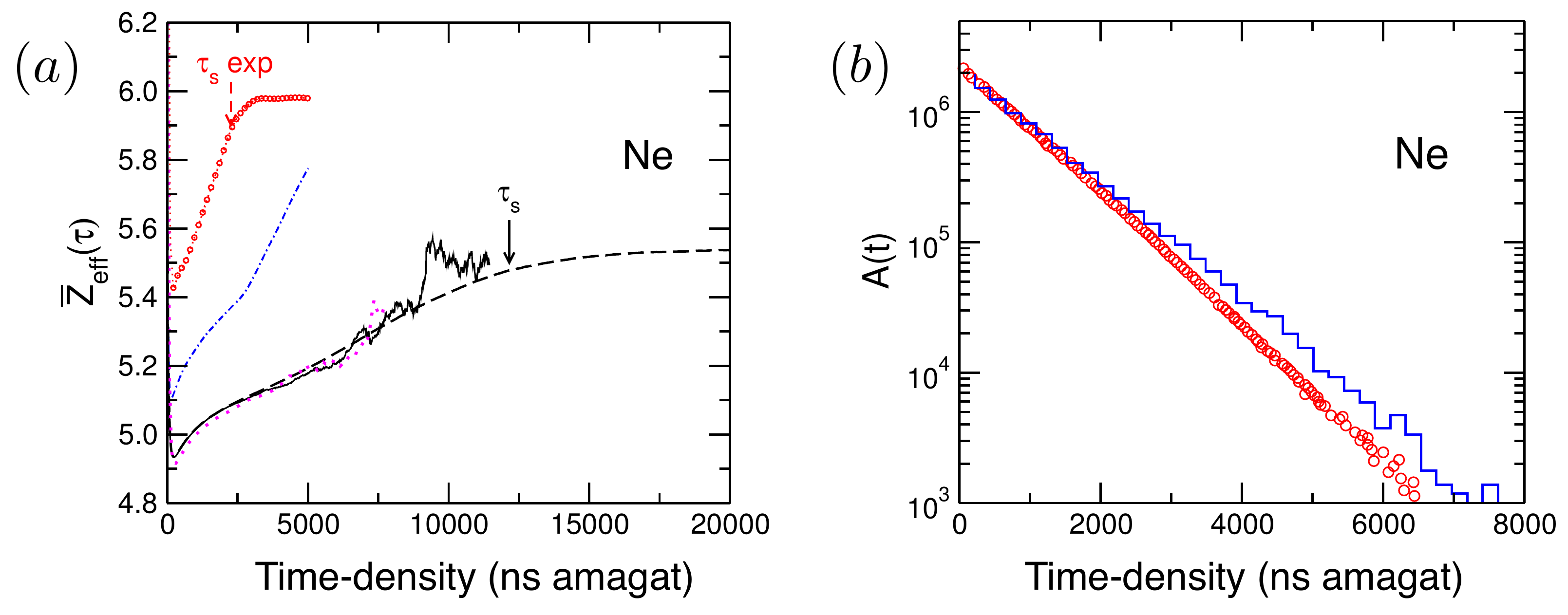}
\caption{(a) $\bar{Z}_{\rm eff}(\tau)$ Ne calculated using $f(k,\tau)$ excluding and including loss of particles due to annihilation, initially distributed uniformly in energy (black dashed and solid lines, respectively), and including annihilation with initial energy equal to the Ps-formation threshold (magenta dotted line). 
Experiment of Coleman \emph{et al.}~\cite{Coleman:1975} (red circles) and FP calculation of Campeanu and Humberston \cite{Campeanu:1977} (blue dashed-dotted lines). Arrows mark the calculated (black) and experimental shoulder lengths (red) $\tau_s$. 
(b) The calculated lifetime spectrum, i.e., observed annihilation rate $A(\tau)=-dF(\tau)/d\tau$, for Ne (blue staircase) (in arbitrary units), compared with experiment \cite{Coleman:1975} (red circles). 
\label{fig:zefft_ne}}
\end{center}
\end{figure*}

Figure \ref{fig:zefft} shows $\bar{Z}_{\rm eff}(\tau)$ for He, Ar, Kr and Xe, obtained in the calculations that excluded or included particle loss due to annihilation, for positrons initially distributed uniformly in energy, and with initial energy equal to the positronium-formation threshold. 
The latter distribution is unphysical, but provides an upper limit on the cooling time.
For all the noble gases the increase in $Z_{\rm eff}(k)$ as $k\to0$ results in the evolution of $\bar{Z}_{\rm eff}(\tau)$ through a transient `shoulder' region resulting from epithermal annihilation at $k<k_{\rm min}$ \cite{Falk:1964,Tao:1964,Paul:1964} towards its steady-state thermal value 
$\bar{Z}_{\rm eff} \equiv \int_0^{\infty} Z_{\rm eff}(k) f_{T}(k) dk$,
where $f_{T}(k)$ is the Maxwell-Boltzmann distribution taken at $T=293$ K.
Comparisons between theory and experiment are focussed around the shoulder,
which is somewhat insensitive to the initial distribution of the positrons \cite{Farazdel:1977,Griffith:1979} as a result of the bunching around the minimum in $B(k)$. 
The traditional measure of the thermalisation time in positron-gas studies is the `shoulder length' $\tau_{\rm s}$, defined via
$\bar{Z}_{\rm eff}(\tau_{\rm s}) \equiv \bar{Z}_{\rm eff} - 0.1\Delta \bar{Z}$ 
where $\Delta \bar{Z} =\bar{Z}_{\rm eff}-\bar{Z}_{\rm min}$, and $\bar{Z}_{\rm min}$ is the minimum of $\bar{Z}_{\rm eff}(\tau)$ \cite{Paul:1968}. 
The calculated $\tau_s$ are given in Table \ref{table:times} along with experimental and previous theoretical results. As noted by Al Qaradawi \emph{et al.}~\cite{AlQaradawi:2000}, and by Paul and Leung \cite{Paul:1968}, systematic errors in experiments may outweigh statistical uncertainties in determining $\tau_{\rm s}$, and the experimental shoulder lengths may be expected to have uncertainties $\sim 10\%$.
We now consider the results for each atom in turn, postponing the discussion of Ne as it is atypical.

\emph{Helium:--}The calculated $Z_{\rm eff}(\tau)$ is seen to be insensitive to both the initial distribution and whether depletion of the distribution is included or not.
It is known that the MBT slightly underestimates the thermal $\bar{Z}_{\rm eff}$ in He, predicting a value of $\bar{Z}_{\rm eff}$=3.79 compared with the experimental value of 3.94 \cite{Coleman:1975}. 
Scaling the calculated $Z_{\rm eff}(\tau)$ to the long-time steady-state experimental value, we find
excellent agreement with experiment around the shoulder region, with the calculated shoulder length $\tau_s = 1839$ ns amg agreeing to within 5\% of the experimental value of $\tau_{\rm s}= 1700\pm50$ ns amg.
The overall shape and length of the calculated shoulder are in better agreement with experiment than the FP calculation of Campeanu and Humberston \cite{Campeanu:1977}, who used the cross sections from their Kohn variational calculations.
The recent diffusion-model calculation of Boyle \emph{et al}.~\cite{Boyle:2014} is also shown. 
It relied on a carefully tuned model polarisation potential and zeroth-order $Z_{\rm eff}(k)$ scaled by enhancement factors, and produced $\tau_{\rm s} = 1618$ ns amagat. Overall, there is good agreement between the present calculation, that of Boyle \emph{et al.}~and experiment. This complements the excellent agreement between the MBT and variational calculations, and measurements of the elastic scattering cross sections \cite{DGG_posnobles}.

\emph{Argon:--}The present calculated $Z_{\rm eff}(\tau)$ are sensitive to the initial distribution at small times, but the overall cooling times for both distributions are similar. The result is weakly sensitive to whether depletion of the distribution is included or not.
The calculated thermal $\bar{Z}_{\rm eff}=26.0$ is in close to the value of 26.77 measured by Coleman \emph{et al.}~\cite{Coleman:1975}. 
The calculated shoulder time $\tau_{\rm s}=369$ ns amg is in excellent agreement with the measured value of 362$\pm5$ ns amg \cite{Coleman:1975}, with reasonable overall agreement in the shape of the shoulder. A much smaller shoulder length was measured in the Al Qaradawi experiment, which was suspected to have suffered from the presence of impurities \cite{AlQaradawi:2000}. 
The present calculations show better agreement with experiment than the FP calculation of Campeanu \cite{Campeanu:1981}, which used the polarised-orbital cross section \cite{arphase}.

\emph{Krypton:--}The calculated shoulder length and $\bar{Z}_{\rm eff}$ are in excellent agreement with experiment of Wright \emph{et al.}~\cite{Wright:1985} (fluctuations at long times are a result of the small number of positrons remaining). The FP calculation of Campeanu \cite{Campeanu:1982}, which used the polarised-orbital cross sections \cite{krxephase}, underestimates $\bar{Z}_{\rm eff}$.

\emph{Xenon:--}The case of Xe is special, because of the strong peaking of $\bar{Z}_{\rm eff}$ at small $k$, which means that annihilation successfully competes with cooling at epithermal positron momenta.
When positron depletion due to annihilation is neglected, the positron cooling is fast ($\tau_{\rm s}\sim 150$ ns amg), and $Z_{\rm eff}$ plateaus at $\sim 450$.
Including depletion brings the shoulder region and shoulder length into excellent agreement with experiment \cite{Wright:1985} (fluctuations are the result of the small fraction of particles remaining). The FP calculation of Campeanu is in serious disagreement with experiment and the MBT result.
It used the polarised-orbital cross sections of McEachran \emph{et al.}~\cite{krxephase}, which predict elastic scattering cross sections and $Z_{\rm eff}(k)$ that are smaller than the MBT calculation and experiment \cite{DGG_posnobles}. 

The present calculations show that $\bar{Z}_{\rm eff}$ is highly sensitive to the loss of particles due to annihilation. The vigorous increase in $Z_{\rm eff}(k)$ as $k\to0$ leads to a quasi-steady-state long-time distribution whose low-momentum component is found to be suppressed relative to the Maxwell-Boltzmann one, and a steady-state annihilation rate $Z_{\rm eff}(\tau\to\infty) \sim 350$ that is significantly reduced from the calculated true thermal $\bar{Z}_{\rm eff} \sim$ 450.  
The present results thus conclusively explain the discrepancy between the gas-cell measurement of $\bar{Z}_{\rm eff}\sim 320$ of Wright \emph{et al.}~\cite{Wright:1985}, 
and the Penning-Malmberg trap measurement $\bar{Z}_{\rm eff}\sim401$ of the Surko group \cite{xenon_zeff_trap}, whose setup ensures positrons are well thermalised. We remark that by adding small amounts of a lighter, low-$Z_{\rm eff}$ gas, e.g., He or H$_2$, to Xe, Wright \emph{et al.}~measured an increase of the pure Xe $\bar{Z}_{\rm eff}\sim 320$ to $\bar{Z}_{\rm eff}$ = 400--450 \cite{Wright:1985}, which is broadly consistent with the present calculated value and the Surko group measurement. The mechanism for such an increase with an admixture of He is, however, unclear, given that momentum transfer is more effective in Xe as $k\to0$ (see Fig.~\ref{fig:kommtxsec}).


\emph{Neon:--}Figure \ref{fig:zefft_ne} shows $Z_{\rm eff}(\tau)$ for Ne. The calculated shoulder time (in this case calculated excluding loss of particles due to annihilation) is $\tau_{\rm s}=12,000$ ns amg, by which time the fraction of positrons remaining is practically zero.
It is drastically longer than the measured value $\tau=2700$ ns amg \cite{Coleman:1975} (note that the FP calculation of Campeanu is also slower and has not reached a steady state value). 
This serious disagreement is in spite of the good agreement of the MBT elastic scattering cross section with experiment, including at the Ramsauer minimum \cite{DGG_posnobles}. 
Moreover, the present calculation is consistent with that expected from mass scaling the He result (which has a similar sized $\sigma_{\rm t}$) $\tau_{\rm th}^{\rm He}M_{\rm Ne}/M_{\rm He}\sim5\tau_{\rm th}^{\rm He}\sim 14,500$ ns amg.

The experimental shoulder length was determined by Coleman \emph{et al.}~\cite{Coleman:1975} via straight line fits to the lifetime spectrum.
However, in spite of the serious discrepancy in $\tau_s$, the calculated and measured lifetime spectra are in surprisingly good agreement [see Fig.~\ref{fig:zefft} (b)]. 
Importantly, the calculated $\tau_{\rm s}=12,000$ ns amg and $\tau_{\rm th}\sim21,000$ ns amg are much longer than the 0--8000 ns amg considered in the experimental analysis.
As seen in Fig.~\ref{fig:annfrac}, at $\tau\lesssim8000$ ns amg the vast majority of positrons have already annihilated, after cooling had effectively stalled around the minimum in $B(k)$. 
Since the $Z_{\rm eff}(k)$ is a reasonably flat function around the minimum, a signal of many to all of the positrons annihilating at a momenta close to the minimum would be observed as a levelling off of $\bar{Z}_{\rm eff}(\tau)$, i.e., a characteristic cooling rate measured over a long time in the lifetime spectra, which could have erroneously been interpreted as the true thermal $\bar{Z}_{\rm eff}$. 
A second possible source of error in the experimental measurement is that it was affected by the presence of impurities. It is known that positron cooling in gases like CO$_2$ and N$_2$O is fast, on a time scale of $\sim$0.1 ns amagat \cite{AlQaradawi:2000}. 
The presence of even minute amounts of impurities could thus lead to a significant reduction in the cooling time.

\section{Discussion\label{sec:summary}}
Positron cooling noble gases has been studied using Monte-Carlo simulations based on accurate cross sections calculated \emph{ab initio} from many-body theory.  
The fraction of positrons surviving annihilation to thermalisation has been shown to be strikingly small. For Xe, the time-varying dimensionless annihilation rate $\bar{Z}_{\rm eff}(\tau)$ is shown to be strongly affected by the depletion of positrons due to annihilation.
The vigorous increase of the positron annihilation rate as $k\to0$ gives rise to a quasi-steady-state distribution whose low-momentum component is suppressed relative to the Maxwell-Boltzman one. This suppression leads to a reduction in the steady-state $\bar{Z}_{\rm eff}$, conclusively explaining the long-standing discrepancy between the gas cell and trap-based measurements of $\bar{Z}_{\rm eff}$. 
Importantly, the current approach has enabled the simultaneous probing of the energy dependence of the scattering cross sections and annihilation rates.
Overall, the use of the accurate MBT data gives the best agreement to date with experiment for all noble gases except Ne, the experiment for which is proffered to have suffered from incomplete knowledge of the fraction of positrons surviving to thermalisation and/or the presence of impurities. New experiments of positron cooling in neon are now warranted. 
Such experiments may include complementary measurements of the time-varying $\gamma$ spectra during cooling, which have been shown to be sensitive to the positron cooling times \cite{DGG_gamcool}. 

\section{Methods}

\subsection{Monte-Carlo simulation of positron cooling and annihilation}
The momentum $k(\tau_i)$ of an individual positron is determined over an equidistant grid in time-density $\{\tau_i\}$ with step size $\Delta \tau$ as follows.
The velocity of a gas particle is sampled from the Maxwell-Boltzmann distribution at 293 K. 
Both it and the positron velocities are transformed to the centre-of-mass frame, in which the energy available for the collision is $E_{\rm CM}=\mu v_r^2/2$, where $\mu$ is the reduced mass and $v_r$ is the relative speed of the positron and gas particle. 
A uniformly distributed random number $r_1=U[0,1]$, is drawn, and a collision is deemed to occur if $r_1<P=W\Delta \tau$, where $W=n_gv_r\sigma_{\rm tot}$ is the probability rate of  annihilation or elastic collision, with $\sigma_{\rm tot}=(\sigma_{\rm el}+\sigma_{\rm a})$, subject to the requirement that $P=W\Delta \tau \ll 1$ (in practice we demand that $P=W\Delta t<0.1$). Here $\sigma_{\rm el}$ is the elastic scattering cross section, which is determined as the integral of the differential elastic cross section
$\varrho={d\sigma}/{d\Omega} = |f(\theta)|^2$, where $f(\theta)$ is the scattering amplitude. The scattering amplitude is calculated from the MBT calculated phase shifts for positron angular momenta up to $\ell=$0, 1 and 2, with all $\ell>2$ approximated using the leading $k^2$ term in the expansion \cite{omalley} [see Sec.~III D of \cite{DGG_posnobles} for details, specifically Eqn.~(31)]. 
(It has been verified in the present work that higher-order corrections \cite{AliFraser} lead to $<3$\% change in $\sigma_{\rm t}$ at the highest positron momentum in Xe).
If a collision occurs, a second random number $r_2=U[0,1]$ is drawn.
If $r_2<\sigma_a/\sigma_{\rm tot}$ the event is deemed to be annihilation and the particle is removed from the simulation, otherwise it is an elastic collision and the positron velocity is updated as follows. The scattering angle $\theta$ is sampled from the differential cross section by finding the root of
$r_3 = {2\pi}{\sigma_{\rm el}^{-1}}\int_0^{\theta} \varrho \sin\theta' d\theta'.$
where $r_3=U[0,1]$.
In the centre-of-mass frame elastic scattering is symmetric with respect to the azimuthal angle $\phi$, which is thus chosen randomly. The momentum distribution $f(k,\tau)$ results from binning positron momenta at each $\tau_i$.

\section{Code availability} The simulations were performed using the Monte-Carlo program {\tt ANTICOOL} developed by the author, which will be detailed in a following article to be submitted to Computer Physics Communications. 
The code can be made available by the author on request.
\section{Data availability} 
The data sets generated during and/or analysed during the current study are available from the corresponding author on request. They will be made available via the Queen's University Belfast data repository on publication.

\section{Acknowledgements}
I thank Gleb Gribakin for insightful discussions, for encouraging my attention to this problem and for useful comments on the manuscript, and Mike Charlton, Christopher Harvey, Matthew Lee and Patrick Mullan for valuable discussions. 
This work was supported by the United Kingdom Engineering and Physical Research Council (EPSRC), Grant EP/N007948/1. 


%

\end{document}